OXFORD | GLOBAL PROJECTS

# Quantitative Cost and Schedule Risk Analysis of Nuclear Waste Storage

10 December 2018

Report prepared by

Dr Alexander Budzier[1]

Prof Bent Flyvbjerg[2]

Andi Garavaglia[3]

Andreas Leed[3]

[1] Director, Oxford, Global Projects;
   Fellow in Management Practice, Saïd Business School, University of Oxford

[2] Director, Oxford Global Projects;
   BT Professor and Chair of Major Programme Management, Saïd Business School, University of Oxford

[3] Researcher, Oxford Global Projects

**EXECUTIVE SUMMARY**

This study was commissioned to provide an independent, outside-in estimate of the cost and schedule risks of nuclear waste storage projects. Nuclear waste storage is here defined as special facilities for the wet and dry storage and disposal of high-level (HILW) and low and intermediate-level nuclear waste (LILW).

The study analyzed cost and schedule risk profiles of past, completed nuclear waste storage (n = 22), nuclear new builds (n = 194) and underground mining projects (n = 31). The study found:

- The cost risk of nuclear waste storage projects is similar to the cost risk of nuclear power projects;
- The cost risk of underground mining is lower than the cost risk of nuclear waste storage projects;
- The schedule risk of nuclear waste storage projects is similar to the schedule risk in other nuclear projects and similar to underground mining projects;
- For cost risk the study found, based on a reference class of 216 past, comparable projects:
  - Cost overrun will be 67% or less, with 50% certainty, i.e. 50% risk of overrun above 67%;
  - Cost overrun will be 202% or less, with 80% certainty, i.e. 20% risk of overrun above 202%;
- For schedule risk the study found, based on a reference class of 200 past, similar projects:
  - Schedule overrun will be 40% or less, with 50% certainty, i.e. 50% risk of schedule overrun above 40%;
  - Schedule overrun will be 104% or less, with 80% certainty, i.e. 20% risk of overrun above 104%.

The Reference Class Forecasting approach employed in this study allows decision makers to identify the needed uplifts based on their risk appetite. The Reference Class Forecast is based on the assumption that the Swiss Nuclear Waste Storage project is no more and no less risky than past, similar, completed projects.

If the decision makers seek a 50% certain cost estimate this study recommends a 67% cost contingency. This entails a total estimate of approximately CHF 14 billion, which has a 50% likelihood of being sufficient and a 50% likelihood of being exceeded. This is higher than the most recent CHF 12 billion estimate by UVEK[4].

If decision makers are more conservative and seek 80% certainty of the cost estimate (P80), this study recommends a 202% cost uplift. This entails a total cost of approximately CHF 25 billion.

---

[4] Eidgenössisches Departement für Umwelt, Verkehr, Energie und Kommunikation (UVEK)

## BACKGROUND TO THIS STUDY

Every five years, the operators of Swiss nuclear power plants are legally obliged to prepare a cost forecast for the decommissioning and disposal costs of the five Swiss reactors.

The most recent cost estimate was prepared in 2016 using a new methodology (KS16)[5]. This is the first time that uncertainty factors have been reported. In addition to the initial costs estimated as overnight costs, it also includes risk mitigation costs, forecast inaccuracies, risks and opportunities. The study also identifies various risks that have not been included in the costs, as the probability of their occurrence was considered low.

The KS16 cost estimate forms the basis for payments by the nuclear power plant operators into state-controlled funds. The anticipated cost is decreed by the authorities, which then form the basis to calculate the operators' payments. The authorities[6] assessed the chargeability of some opportunities differently from that calculated by the authors of the KS16 study and adjusted the total anticipated costs slightly upwards.

The methodology of the 2016 cost study has been consistently assessed as good ("best practice"). This is one of the reasons why a flat-rate safety surcharge of 30%, which was previously prescribed on account of the great inaccuracies and high cost increases of previous cost studies, is suggested to be removed from the relevant regulation in 2018/19.

The construction of the deep repository for radioactive waste has suffered a considerable delay. There is no sufficient worldwide reference for the planned concept.

As a result, the true cost will only be known decades after the last nuclear power plant has been decommissioned - at a point when the owners will most likely have ceased to exist in their present form. The owners will therefore not be able to make any additional payments. Current legislation transfers these cost risks to the general public.

## QUESTIONS FOR THIS STUDY

Based on reviews of the 2016 cost estimate, the Schweizerische Energie-Stiftung has previously called for an increase of the 30% surcharge to 100% until the cost increases have stabilized.

To independently assure this recommendation the Schweizerische Energie-Stiftung commissioned Oxford Global Projects to conduct a study into Cost and Schedule Uplifts.

---

[5] Kostenstudie 2016 (KS16) Schätzung der Entsorgungskosten Zwischenlagerung, Transporte, Behälter und Wiederaufarbeitung, swissnuclear, Fachgruppe Kernenergie der swisselectric, Olten

[6] Eidgenössischen Departement für Umwelt, Verkehr, Energie und Kommunikation (UVEK)

The Schweizerische Energie-Stiftung posed three questions for this study:

- How accurate can a cost study in 2016 really be for Nagra's[7] planned major deep geological repository project? What minimum and maximum cost overruns are to be expected based on experience with other major projects or comparable projects?

- Building and opening the deep geological repository has already been postponed many times in the past. What delays and associated additional costs can be expected based on experience with other major or comparable projects?

- Is the methodology chosen in the 2016 cost study, including a planned optimism bias supplement of 12.5%, adequate in terms of anticipating the actual costs with sufficient probability (e.g. 80%)? Or is it necessary to continue to have an additional flat-rate safety surcharge to cover possible and probable cost overruns?

---

[7] Nationale Genossenschaft für die Lagerung radioaktiver Abfälle (National Cooperative for the Disposal of Radioactive Waste)

## DATA ANALYSIS STRATEGY

Accurate estimates and thus higher-quality project decisions combine what has been called the *"outside view"* and the use of *all the distributional information* that is available. This may be considered the single most important piece of advice regarding how to increase accuracy in forecasting through improved methods, according to Kahneman (2011: 251).

Reference Class Forecasting is a method for systematically taking an outside view on planned actions. Reference class forecasting places particular emphasis on *all* relevant distributional information because such information is crucial to the production of accurate forecasts.

Reference Class Forecasting makes explicit, empirically based adjustments to estimates. In order to be accurate, these adjustments should be based on data from past projects or similar projects elsewhere and adjusted for the unique characteristics of the project in hand.

Reference Class Forecasting follows three steps:

1) Identify a sample of past, similar projects;
2) Establish the risk of the variable in question based on these projects – e.g. identify the cost and schedule overruns of these projects; and
3) Adjust the current estimate – through an uplift or by asking whether the project at hand is more or less risky than projects in the reference class, resulting in an adjusted uplift.

It should be noted that any adjustments to the uplift in the final step ought to be based on hard evidence in order to avoid reintroducing optimism back into the estimate.

Because Reference Class Forecasts are based on the actual outcomes of similar past projects, the method estimates not only the known unknowns of a project, i.e. risks identified ex-ante, but also the unknown unknowns for the project, i.e. risks that have not been identified but may nevertheless impact the project.

## STEP 1 – IDENTIFY A SAMPLE OF PAST, SIMILAR PROJECTS

For this analysis we identified project categories and projects, which may have a risk profile similar to nuclear waste storage facilities.

### Nuclear Waste Storage Projects

The study investigated 53 low and intermediate-level storage projects, 24 high-level storage projects, 10 research facilities. The projects under consideration are listed in Appendix A.

We collected the data from publicly available sources, e.g. annual reports, project updates, investment decision documents. We also contacted the owner-operators to provide project relevant information.

### Nuclear Power Projects

We analyzed 199 nuclear new-build projects from the academic database of Oxford Global Projects. For these projects we were initially considering first-generation plants based on the rationale to identify first-mover projects. However, the analysis showed that risks in nuclear power projects have not decreased over time (see Figure 1). Thus, all these projects were included.

**Figure 1 Cost overruns over time in nuclear new build projects**

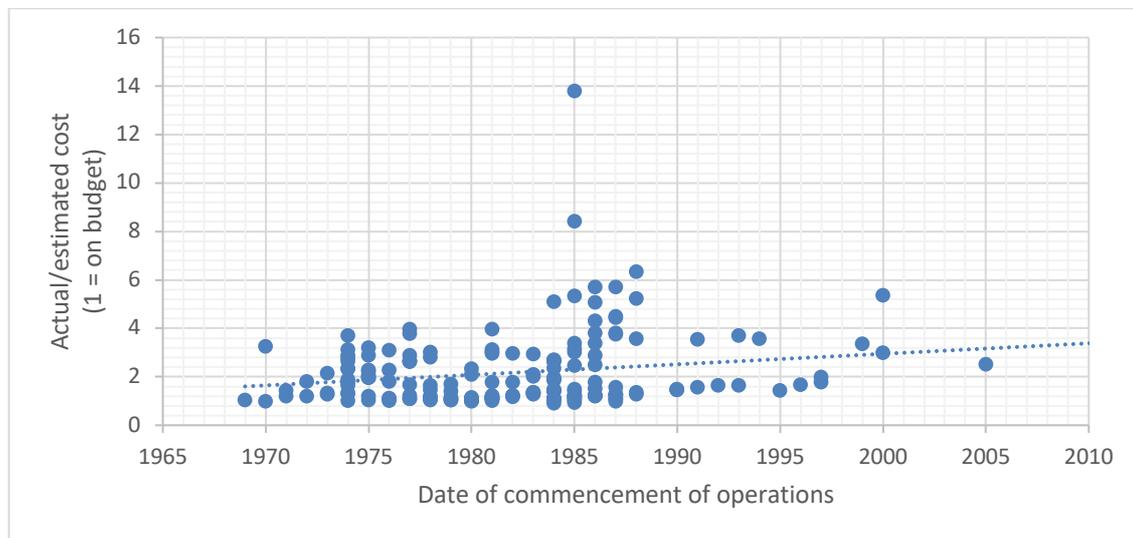

### Underground Mining Projects

Based on the Oxford Global Projects database of 958 mining projects, we identified all underground mining projects exceeding 100m depth and all deep-underground mining projects (>300m depth). In total this provided 31 projects, 16 of these were deep underground mining projects.

**Other Project Types**

We further considered including data on nuclear power plant decommissioning. However, we found sufficient data on nuclear waste storage projects and excluded decommissioning projects from the scope of the study.

The key rationale behind this decision was that based on the waste storage projects a robust reference class could be constructed. Readily available data on nuclear new-builds and underground mining allows a comparison with waste storage projects. Thus, this analysis will be able to assess whether nuclear waste storage is similar, in terms of risk, to other nuclear projects or to underground mining projects.

Nevertheless, studying the cost and schedule risk of nuclear decommissioning could be a fruitful further endeavor to strengthen the analysis and learn from the lessons of these projects.

**Calculating cost overrun and schedule overrun**

Cost overrun is defined as cost overrun = actual cost / estimated cost – 1.
Schedule overrun is defined as schedule overrun = actual cost / estimated cost – 1.

Estimated cost and schedule were baselined at the time of the decision to build (final business case; final investment decision). Actual cost was measured at the time the project was completed.

For the nuclear waste storage facilities, in 3 instances final cost was not yet known, because the project was not fully completed. In these cases, if the project was at least 75% complete, then the cost estimate at this stage was treated as a proxy for actual cost. Previous analysis of construction projects showed that updates to cost and schedule forecasts at a 70-80% of completion are typically reliable because at this stage most risks are known to planners. However, this might introduce a conservative bias into the data, i.e. a forecast based on these data is too low and cost and schedule risk could be higher.

**Testing for comparable risk profiles**

The key advice for a reference class forecast is to base the analysis on as much data as possible (Kahneman 1979b) and not throw out any relevant information. Thus, the final important question in selecting the reference class is whether the different project types have comparable risk profiles.

Table 1 reports the median cost and schedule overruns. The median cost overrun for nuclear new builds and nuclear waste storage projects are in a similar region between 43-73%.

**Table 1 Median cost and schedule overruns in the data**

|  | Cost overrun (%) | | Schedule overrun (%) | |
|---|---|---|---|---|
|  | Sample size (N) | Median | Sample size (N) | Median |
| Underground mines (>100m) | *11* | 33% | *8* | 50% |
| Deep underground mines (>300m) | *20* | 0% | *15* | 50% |
| Nuclear power, new build | *194* | 68% | *177* | 40% |
| Nuclear storage (HLW) | *11* | 40% | *14* | 41% |
| Nuclear storage (LILW) | *11* | 73% | *9* | 40% |

Table 2 tests whether the data on nuclear projects should be pooled across new builds, high-level and low and intermediate-level storage projects. The non-parametric tests found no statistically significant differences between any of the nuclear project types. These results indicate that the medians are not statistically different at any level smaller than 48%, which is HLW compared to LILW.

**Table 2 Non-parametric Wilcoxon test of the cost and schedule overruns of nuclear (p-values, statistically significant differences at p < 0.05 are highlighted in bold)**

| **Cost overrun** | *Deep underground mines* | *Nuclear, storage HLW* | *Nuclear storage, LILW* | *Nuclear power, new-build* |
|---|---|---|---|---|
| *Nuclear, storage HLW* | **0.012** | - | - | - |
| *Nuclear storage, LILW* | **0.033** | 0.519 | - | - |
| *Nuclear power, new-build* | **0.000** | 0.886 | 0.640 | - |
| *Underground mines* | 0.119 | 0.340 | 0.131 | **0.047** |

| **Schedule overrun** | *Deep underground mines* | *Nuclear, storage HLW* | *Nuclear storage, LILW* | *Nuclear power, new-build* |
|---|---|---|---|---|
| *Nuclear, storage HLW* | 0.810 | - | - | - |
| *Nuclear storage, LILW* | 0.740 | 0.820 | - | - |
| *Nuclear power, new-build* | 0.480 | 0.800 | 0.430 | - |
| *Underground mines* | 0.950 | 0.970 | 0.660 | 0.870 |

When the data of nuclear projects are pooled, the difference to underground mining projects is statistically significant for cost overrun ($p < 0.001$, non-parametric test) but not for schedule overrun ($p = 0.570$).

Thus, no sufficiently strong evidence exists that indicates that the data on nuclear should be split into subgroups. In effect, the cost and schedule risk profiles of these projects are statistically similar.

With regards to the mining projects the cost risk profile is statistically significantly different and these should not be considered in the same reference class. With regards to schedule risk they could be combined – no strong statistical evidence suggests that the schedule risk profile is different – however to keep both reference classes consistent they were excluded from the schedule risk analysis below.

**STEP 2 – ESTABLISH THE RISK OF THE VARIABLE IN QUESTION**

The variable in question here is cost and schedule overrun. The available data were sorted from smallest to largest overrun and the cumulative frequency is calculated.

The distribution of cost overrun (Figure 2) shows that nuclear projects do indeed follow a similar cost risk profile to nuclear storage projects up to 300% overrun. Approximately 20% of nuclear projects had a cost overrun above 200-300%; 80% of projects completed equal to or below this level. Figure 3 zooms in on the range of cost overrun up to 300%, which excludes the considerable tail risk that exists in nuclear projects.

**Figure 2 Cumulative frequency of cost and schedule overrun observed in the data**

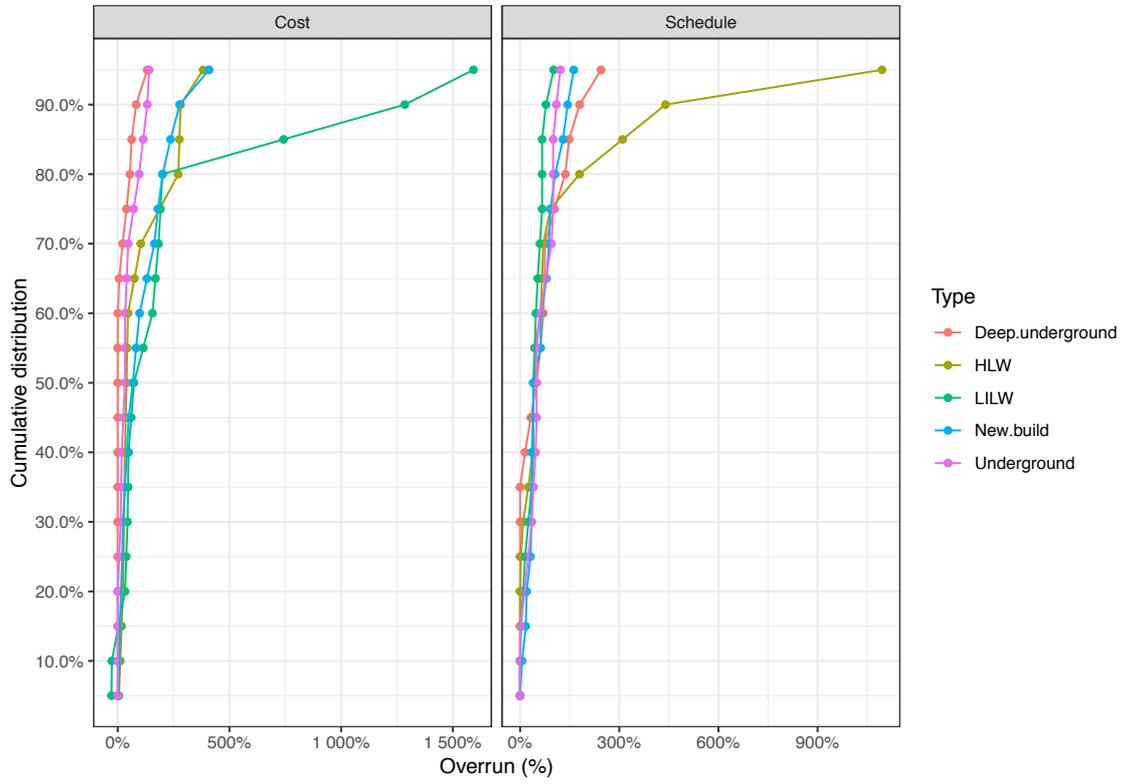

**Figure 3 Cumulative frequency of cost and schedule overrun observed in the data (detail)**

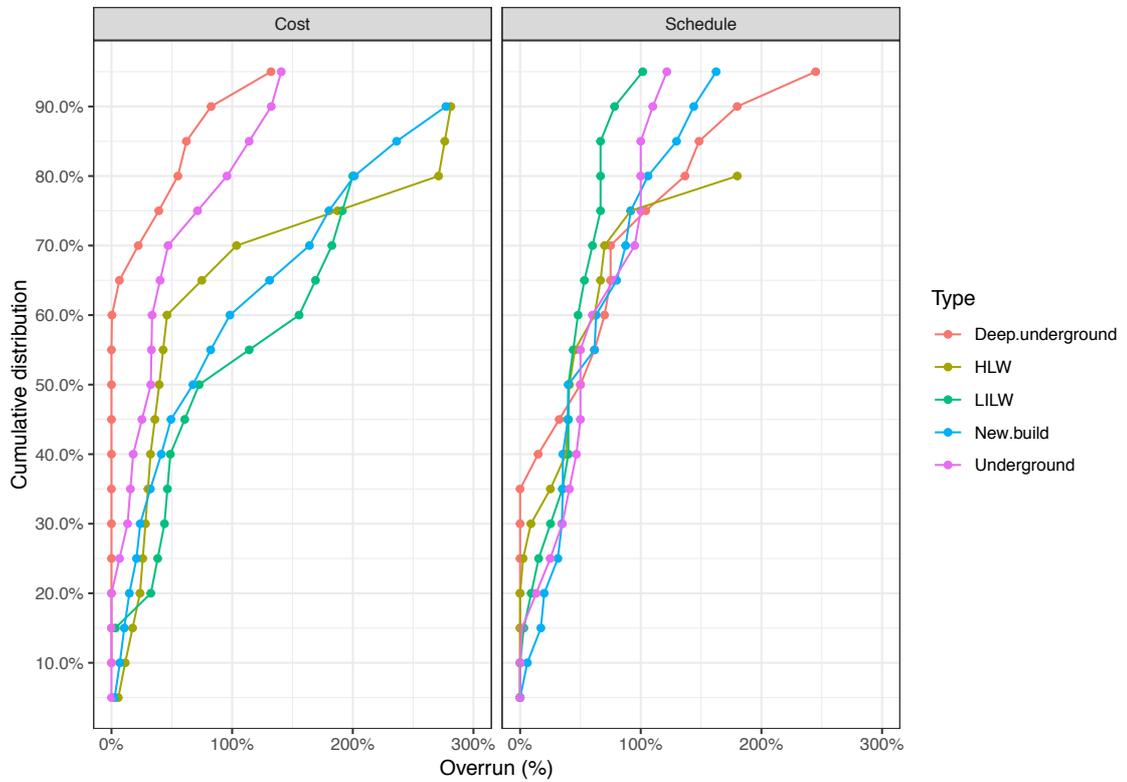

The distribution of schedule overrun shows similar patterns. 20% of projects had a schedule overrun above 100%; 80% of projects completed equal to or below this level. The upper tail of schedule overrun, where large delays to the projects are found, is less fat than the tail of the cost overrun distributions.

The analysis in step 1 found that the nuclear data should be pooled. Figure 4 presents the final reference classes after pooling the data for nuclear projects and mining projects.

**Figure 4 Cumulative frequency of cost and schedule overrun observed in the reference classes (detail)**

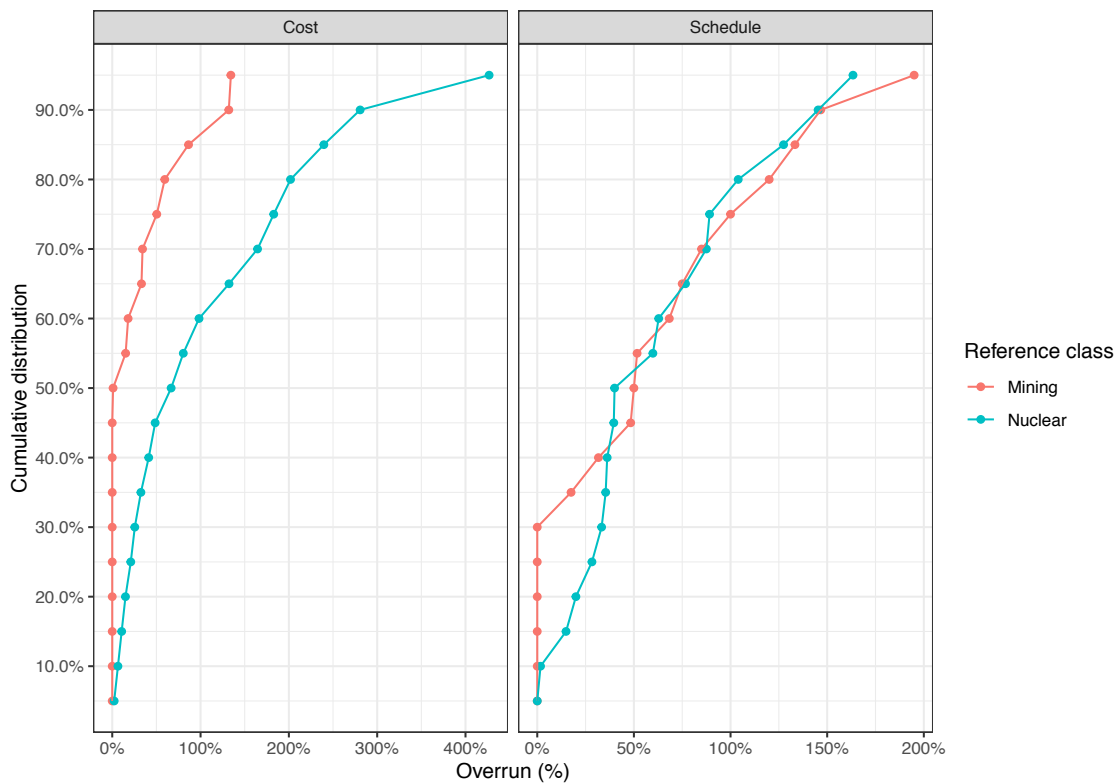

## STEP 3 – ADJUST THE CURRENT ESTIMATE

If the project is no more and no less risky than similar past projects, the reference class forecast directly provides the uplift necessary to de-bias the estimation of risk.

Figure 5 explains how the uplifts are identified from the distribution of overrun.

**Figure 5 Conceptual diagram to identify the required uplifts given decision makers' acceptable chance of overrun, here for cost overrun**

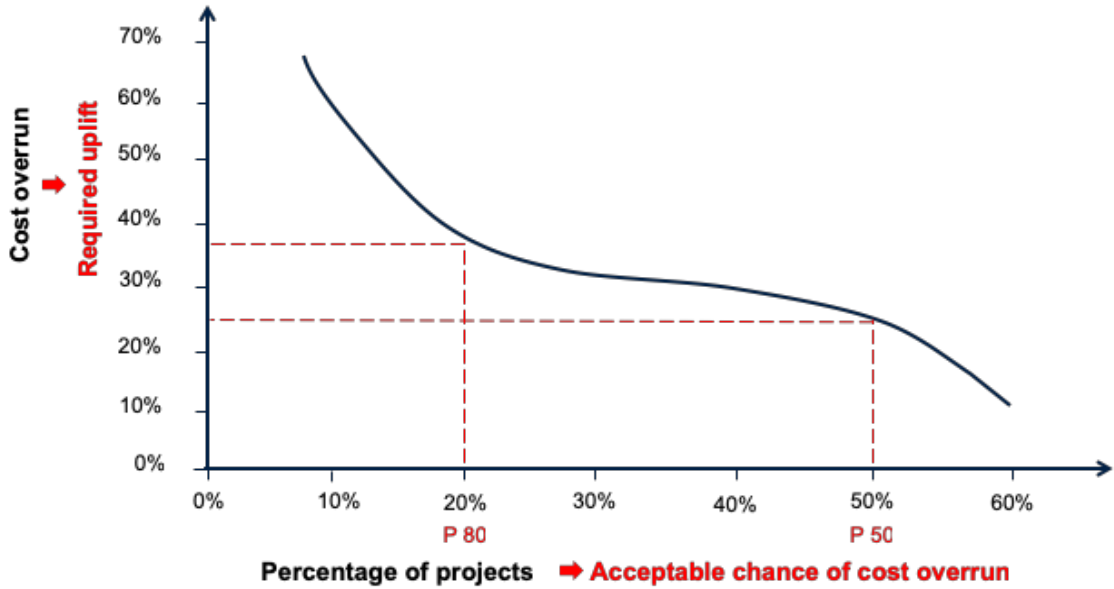

The necessary uplifts are identified by the inverse of the distribution in Figure 2. Figure 5 shows the results, where the percentage of projects becomes the acceptable chance of overrun. The observed overrun becomes the uplift necessary.

**Figure 6 Cost and schedule uplifts to be applied to a project based on the desired level of certainty of decision makers (31 mining projects, 216 nuclear projects for cost; 23 mining projects, 200 nuclear projects for schedule)**

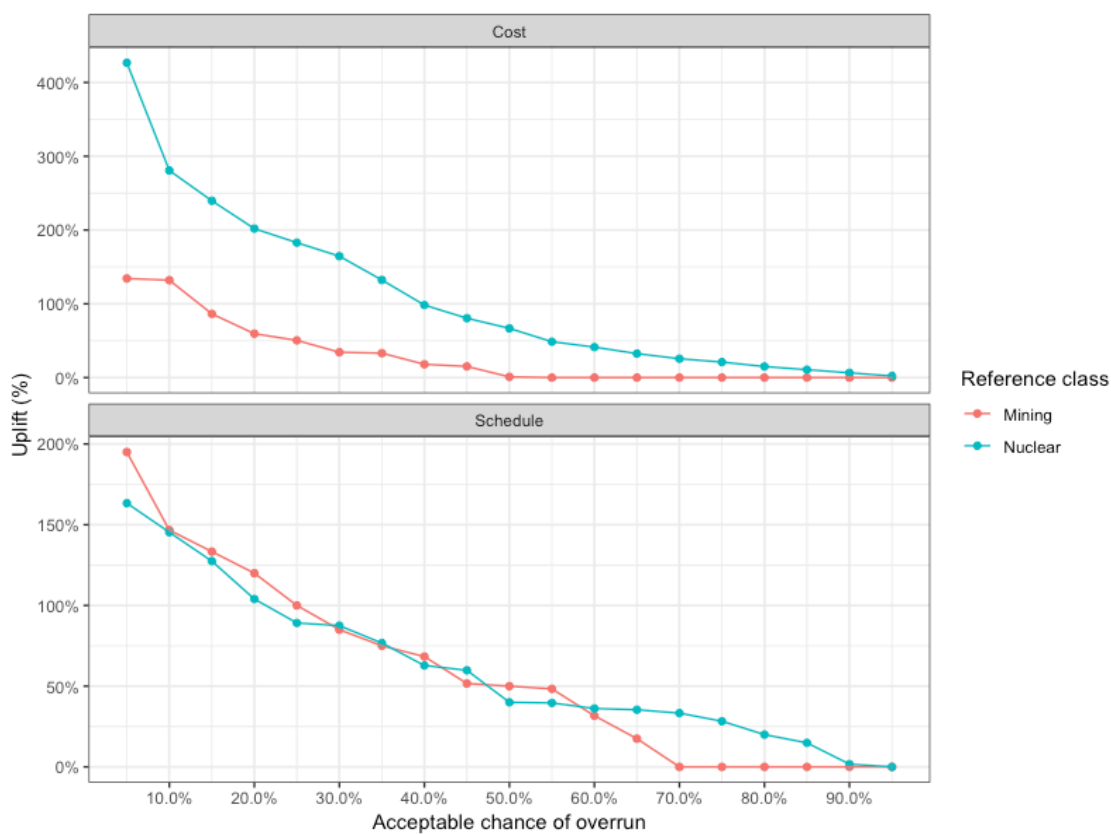

Figure 6 allows decision makers to choose the required cost and schedule uplifts based on their risk appetite. Uplifts are identified by choosing the accepted chance of overrun. The uplifts in Figure 6 are also shown in Table 3 below.

For example, if decision makers are willing to accept a 50% chance of overrun (i.e. they seek a 50% certain estimate or P50 estimate) the required cost uplift in nuclear projects is 67%. Thus, if forecasted cost is uplifted by 67% the new budget will be met with a probability of 50% and exceeded with a probability of 50%, assuming that the proposed project is no more and no less risky than past, similar projects.

The required uplift for a 50% acceptable chance of schedule overrun in nuclear projects is 40%. Thus, if a 40% schedule buffer is added to the project plan then the new finish date will be met with a probability of 50% and exceeded with a probability of 50%, assuming that the proposed project is no more and no less risky than past, similar projects.

The P50 estimate is often used to forecast the most likely cost of projects. However, for big, one-off capital investment projects, decision makers will typically regard a level of 50% certainty to be too low. In this case, decision makers would typically accept a lower chance of overrun, often 20%, i.e. estimates with 80% certainty (P80). A 20% acceptable chance of cost overrun for nuclear projects requires an uplift of 202%, as seen in Figure 6. In this more risk averse scenario, decision makers would have to apply a 202% uplift to their project proposal to ensure that the probability of a budget overrun is reduced to 20%. The schedule uplift at a 20% acceptable chance of overrun is 104%.

In some cases, decision makers have asked for even lower chances of overrun than 20%, for instance 5% for UK's High Speed 2 railway.

**Table 3 Reference Class Forecast for cost and schedule uplifts for nuclear projects, based on the final business case at the decision to build (n = 216 for cost; n = 200 for schedule; uplifts given in percent to be added to the cost and schedule estimate)**

| Acceptable chance of overrun | Level of certainty of the estimate (P-value) | Cost uplift | Schedule uplift |
|---|---|---|---|
| 95% | 5% | 2% | 0% |
| 90% | 10% | 6% | 2% |
| 85% | 15% | 11% | 15% |
| 80% | 20% | 15% | 20% |
| 75% | 25% | 21% | 28% |
| 70% | 30% | 26% | 33% |
| 65% | 35% | 32% | 35% |
| 60% | 40% | 41% | 36% |
| 55% | 45% | 49% | 40% |
| 50% | 50% | 67% | 40% |
| 45% | 55% | 80% | 60% |
| 40% | 60% | 98% | 63% |
| 35% | 65% | 132% | 77% |
| 30% | 70% | 165% | 88% |
| 25% | 75% | 183% | 89% |
| 20% | 80% | 202% | 104% |
| 15% | 85% | 240% | 127% |
| 10% | 90% | 281% | 145% |
| 5% | 95% | 427% | 163% |

## SUMMARY OF FINDINGS

The analysis found:
- The cost risk of nuclear waste storage projects is similar to the cost risk of nuclear power projects;
- The cost risk of underground mining is lower than the cost risk of nuclear waste storage projects;
- The schedule risk of nuclear waste storage projects is similar to the schedule risk in other nuclear projects and similar to underground mining projects;

- For cost risk the study found, based on a reference class of 216 past, comparable projects:
  - Cost overrun will be 67% or less, with 50% certainty, i.e. 50% acceptable chance of overrun above 67%;
  - Cost overrun will be 202% or less, with 80% certainty, i.e. 20% acceptable chance of overrun above 202%;
- For schedule risk the study found, based on a reference class of 200 past, similar projects:
  - Schedule overrun will be 40% or less, with 50% certainty, i.e. 50% acceptable chance of schedule overrun above 40%;
  - Schedule overrun will be 104% or less, with 80% certainty, i.e. 20% acceptable chance of overrun above 104%.

This report was commissioned to consider three specific questions:

1. How accurate can the KS16 cost study really be for Nagra's planned major deep geological repository project? What minimum and maximum cost overruns are to be expected based on experience with other major projects or comparable projects?

2. Building and opening the deep geological repository has been postponed several times. What delays and associated additional costs can be expected based on experience with comparable projects?

3. Is the methodology chosen in the KS16 cost study, which included a planned optimism bias supplement of 12.5%, adequate in terms of anticipating actual costs at the required level of certainty (e.g. 80%)? Or is it necessary to continue to have an additional flat-rate safety surcharge to cover possible and probable cost overruns?

**Accuracy of the KS16 cost study**

The analysis is based on a large sample of statistically comparable projects. The analysis found that the profile of cost and schedule overrun in nuclear storage projects is not statistically different between high-level (HLW), low and intermediate-level waste (LILW) projects.

The analysis also found that nuclear new builds have a similar profile of cost and schedule overruns to nuclear storage projects. On the other hand, mining projects had a statistically significantly different profile of cost overrun.

This seems to indicate that the causes of cost and schedule overrun in nuclear storage projects are more similar to the causes of overrun in other nuclear projects; and different from the causes of overrun in underground mining projects.

In the historic data, cost overruns range from -30% to 1900%, with the middle 90% of projects ranging from 2% to 427%. The cost uplift required at a 20% acceptable chance of overrun is 202%. The cost uplift required at a 50% acceptable chance of overruns is 67%.

The KS16, UVEK and STENFO recommended contingencies between 23% and 46%. The historic data in our analysis suggest that the recommended contingencies are underestimating the true risks of the project and therefore insufficient to cover the likely cost overrun.

**Schedule risk**

In the historic data, delays range from -15% to 2200%. The middle 90% of projects had delays ranging from 0% to 163%. The schedule uplift required at a 20% acceptable chance of overrun is 104%. The schedule uplift required at a 50% acceptable chance of overrun is 40%.

The data show a strong correlation between schedule overrun and cost overrun ($r = 0.39$, $p < 0.001$).

A robust regression (fitting is done by iterated re-weighted least squares to limit the effect of the large outliers in the tail of overrun) indicates that for every 10% of schedule delay cost overrun increases by 11% (with 95% confidence a 10% schedule delay added 9%-13% cost overrun).

According to the KS16 cost study the final decision to build is expected in 2031 with construction of the Swiss nuclear storage planned from 2032-2049 for the LILW and 2060 for the HLW facilities, i.e. the time between final decision and readiness for operations is 30 years.

Based on the CHF 8.229 billion cost estimate a 1-month delay increases the average cost overrun by CHF 24 million. Every day of delay increases the average cost overrun by CHF 0.8 million.

**Adequacy of current risk contingency**

The historic data, used in this study, were based on the cost estimate at the decision to build. For some projects, this estimate possibly includes some level of contingency. However, it was impossible to establish the size of this possible contingency.

The KS16 cost study (see Table 4 below) estimated the needed cost contingency for the future work at 23.2%. To give the KS16 contingency estimate the benefit of doubt, i.e. assuming that the historic data did not include any contingencies for cost, then a 23% contingency would have been sufficient for 28% of past, similar projects. 72% of projects exceeded this contingency.

Thus, if the deep geological repository faces similar risks as past, comparable projects and if the project is no more and less risky than other projects, then the risk estimate has a certainty of nearly 30%[8].

---

[8] In our data collection we were not able to reconstruct the level of contingency included in the cost forecasts at the decision to build. However, cost estimation guidelines for nuclear power plant decommissioning by the IAEA (Report No. NW-T-2.4) suggests that 10% contingency is commonly applied.

## Comparison of the size of contingencies with other studies

Various institutions have made recommendations about the required cost contingencies for the project. Table 4 compares these to the findings of this study.

**Table 4 Comparison of recommended cost contingencies and resulting cost estimates (CHF billions, excluding incurred cost)**

|  | Total contingencies recommended | Cost for 2 storage facilities (Kostenstudie 2016) | Cost for 2 storage facilities (STENFO) | Total cost for 2 facilities (UVEK) |
|---|---|---|---|---|
| **Base cost** |  | 8.13 | 8.23 | 12.38 |
| swissnuclear | 23.20% | 10.01 | 10.14 |  |
| STENFO | 38.50% |  | 11.40 |  |
| UVEK | 50.50% |  | 12.38 |  |
| SES | 100.00% |  |  | 24.10 |
| **This study (most likely, P50)** | 67.00% | 13.57 | 13.74 |  |
| **This study (conservative, P80)** | 202.00% | 24.54 | 24.85 |  |

Sources:
swissnuclear/STENFO: Summary_Überprüfung KS16_final, p.11
UVEK: https://www.admin.ch/gov/de/start/dokumentation/medienmitteilungen.msg-id-70407.html
SES: https://www.energiestiftung.ch/medienmitteilung/sicherheitszuschlag-schuetzt-steuerzahlende-am-besten.html

The approach in this study allows decision makers to identify the needed uplifts based on their risk appetite. If the decision makers seek a 50% certain cost estimate, then this study recommends a 67% contingency. This translates into a total cost estimate of approximately CHF 14 billion, which then has a 50% likelihood of being sufficient and a 50% likelihood of being exceeded. This is higher than the CHF 12-billion estimate by UVEK.

If decision makers are more conservative and seek 80% certainty of the estimate (P80), then this study recommends a 202% cost uplift. This translates into a total cost of approximately CHF 25 billion.

If we assume that the projects in the reference class included a 10% contingency then the 23% contingency in the Swiss project has a certainty of approximately 17% (P17).
If we assume that the projects in the reference class included a contingency similar to the one planned in the Swiss project (contingency of approximately 20%) then the certainty of the Swiss estimate is only 3% (P03), because 97% of similar, past projects exceeded their budget.

# APPENDIX A

The projects we investigated were:

- **High-level waste research facilities**
  - HADES Underground Research Facility
  - AECL Underground Research Lab
  - Meuse/Haute Marne Underground Research Laboratory
  - Horonobe Underground Research Lab
  - Mizunami Underground Research Lab
  - Korea Underground Research Tunnel
  - Aspo Hard Rock Lab
  - Grimsel Test Site
  - Mont Terri Rock Laboratory
  - Yucca Mountain Storage Facility
- **High-level waste projects**
  - Belgoprocess- Building 136
  - Eurochemic
  - Tihange
  - DGR Canada
  - Darlington
  - ISFSF
  - Onkalo
  - La Hague Vitrification Facility (R7)
  - Marcoule (UP1)
  - Cigeo
  - Ahaus
  - Gorleben Salt Dome
  - Recyclable Fuel Storage Centre
  - Rokkasho-mura
  - HABOG
  - Individual Interim Storage Facility
  - ATC
  - DGR Spain
  - Forsmark Area Repository
  - CLAB
  - Central Interim Storage Facility (ZZL)
  - THORP Receipt & Storage
  - Vitrified HLW Storage Facility
  - Yucca Mountain Storage Facility
- **Low-level and intermediate-level storage projects**
  - cAt Project
  - Belgoprocess- Building 151
  - Permanent Repository for Radioactive Waste (PRRAW)
  - NRRAW
  - Bruce Site
  - Richard Repository
  - Bratrstvi
  - Dukovany
  - VLJ
  - Loviisa
  - Centre de l'Aube
  - Schacht Konrad
  - National Radioactive Waste Repository
  - RWTDF
  - Avogadro Repository
  - LLW Disposal Center
  - TBD
  - Stabatiske
  - Maisiagala
  - COVRA
  - Rozan
  - ITN, DPRSN
  - WLDC
  - Baita Bihor
  - Saligny
  - National RAW Repository
  - JAVYS
  - Brinje
  - Vrbine
  - El Cabril LLW & ILW Disposal Facility
  - El Cabril II
  - SFR Final Repository
  - Beznau I
  - Beznau II
  - Muhlenberg
  - Gosgen NPP
  - Leibstadt
  - General Storage Facility
  - National Collection Centre & Federal Storage Facility
  - Drigg
  - Chapelcross
  - Harwell
  - Hinkley Point A
  - Berkeley
  - Bradwell
  - WIIP
  - Federal Waste Facility
  - Texas Compact Facility
  - Barnwell
  - Clive
  - Oak Ridge
  - Richland
  - Vaalputs
- **Nuclear power plants**
  - Shoreham
  - Clinton
  - Diablo Canyon 1
  - Fermi 2
  - Clinton
  - River Bend 1
  - Rajasthan Atomic Power Station III and IV
  - Waterford 3
  - Nine Mile Point 2
  - WPPSS 2
  - Diablo Canyon 2
  - Pickering A Unit 1
  - Beaver Valley 2
  - Shearon Harris 1
  - Limerick 1
  - Salem 2
  - Salem 1
  - Millstone 3
  - Perry 1
  - Joseph M. Farley 1
  - Byron 2
  - Darlington
  - Catawba 1
  - Braidwood
  - Kakrapar I and II
  - Narora Atomic Power Station I and II
  - Y-12 Facility
  - Byron 1
  - Palo Verde 1
  - Civaux 2
  - Tsuruga-1

- Calvert Cliffs 1
- Grand Gulf 1
- McGuire 1
- Three Mile Island 1
- St. Lucie 1
- Joseph M. Farley 2
- Susquehanna 2
- North Anna 1
- Kaiga I and II
- Sequoyah 1
- Sequoyah 2
- San Onofre 2
- Browns Ferry 3
- Browns Ferry 2
- Browns Ferry 1
- Hope Creek
- Olkiluoto Nuclear Power Plant 3
- Davis-Besse 1
- Cooper
- Kalininskaya
- Pickering A Unit 4
- Virgil Summer 1
- McGuire 2
- LaSalle 1
- Calvert Cliffs 2
- Peach Bottom 2
- Brunswick 2
- Callaway
- Crystal River 3
- Tarapur III and IV
- Palo Verde 2
- Wolf Creek 1
- Pickering B
- LaSalle 2
- Arkansas Nuclear 2
- Fort Calhoun 1
- Shimane-1
- Beaver Valley 1
- Edwin I. Hatch 1
- Rancho Seco
- Rajasthan Atomic Power Station I
- Duane Arnold
- St. Lucie 2
- North Anna 2
- Susquehanna 1
- Civaux 1
- Donald C. Cook 1
- Millstone 2
- San Onofre 3
- Trojan
- Bruce A (1969-1978)
- Bruce A Units 3 & 4 (2001-2004)
- Three Mile Island 2
- Madras Atomic Power Station I
- Kewaunee
- Leibstadt
- Surry 1
- Indian Point
- Ikata 2
- Madras Atomic Power Station II
- Chooz B2
- Rajasthan Atomic Power Station II
- Zion 2
- Point Lepreau Nuclear Generation Station Refurbishment
- Ohi-1
- Chooz B1
- Brunswick 1
- Arkansas Nuclear 1
- Penly 2
- Golfech 2
- Fukushima 1-5
- Chinon B4
- Cattenom 4
- Bruce B
- Chinon B3
- Penly 1
- Gravelines 6
- Golfech 1
- Fukushima 1-4
- Cattenom 3
- Gravelines 5
- Cruas 4
- Sizewell B
- Palisades
- Cruas 2
- Cruas 3
- Edwin I. Hatch 2
- Pickering A
- Takahama-3
- Nogent 2
- Blayais 3
- Blayais 4
- Cruas 1
- Surry 2
- Peach Bottom 3
- Takahama-1
- Belleville 2
- Zion 1
- Nogent 1
- Chinon B2
- Belleville 1
- Cattenom 2
- Kashiwazai-Kariwa-1
- Chinon B1
- Flamanville 2
- Fessenheim 2
- Cattenom 1
- Paluel 4
- Blayais 2
- Dampierre 4
- Takahania-2
- Flamanville 1
- St Alban 2
- Fukushima 1-1
- Hamaoka-2
- Mihama-2
- Fukushima II-1
- Paluel 3
- Bruce A Units 1 & 2
- St Alban 1
- Dampierre 3
- Blayais 1
- Bugey 2
- Dampierre 2
- Fukushima 1-6
- Harnaoka-1
- Gravelines 4
- Sendai-1
- Ohi-2
- Paluel 2
- Fessenheim 1
- St Laurent B2
- Ikata 1
- Tsuruga-2
- Paluel 1
- Dampierre 1
- Fukushima 1-3
- Hamaoka-3
- Gravelines 3
- Tricastin 4
- St Laurent B1
- Bugey 3
- Tricastin 3
- Takahama-4
- Bugey 4
- Bugey 5
- Fukushima II-3

- Tarapur I and II
- Gravelines 2
- Tokai-Daini
- Genkai-1
- Onagawa-1
-

- Genkai-2
- Tricastin 1
- Tricastin 2
- Mihama-3
- Fukushima 1-2

- Gravelines 1
- Mihama-1
- Fukushima II-4
- Sendai-2
- Fukushima II-2

# APPENDIX B – THEORY BEHIND OVERRUN

Owner-operators and builders of projects tend to explain cost and schedule overruns in major projects as a result of unforeseen ground conditions, project complexity, scope and design changes, weather, delays in site access and possession, delays in obtaining permits etc. (see Cunningham 2017, for a review of studies of causes of cost and schedule overruns).

No doubt, all of these factors at one time or another contribute to cost overrun and schedule delay, but it may be argued that they are not the real, or root, cause. The root cause of overrun is the fact that project planners tend to systematically underestimate or even ignore risks of complexity, scope changes, etc. during project development and decision making.

The root cause of cost overrun and schedule delay is not that unforeseen conditions and adverse events happen to a project. The root cause is found in what a project did or did not do to prepare for unforeseen conditions and adverse events.

**Root causes of cost overruns and schedule delays**

Most projects change in scope during progress from idea into reality. Changes may be due to uncertainty regarding the level of ambition, the exact corridor, the technical standards, safety, environment, project interfaces, geotechnical conditions, etc. In addition, the prices and quantities of project components are subject to uncertainty.

Hence, some degree of cost and schedule *risk* will always exist. Such risk is however not unknown and should be duly estimated and reflected in the project documentation at any given stage. Hence, cost overruns and schedule delays should be viewed as *underestimation* of cost and schedule risk.

Only identifying the root causes of what causes projects to underestimate cost and schedule risk allows planners and decision makers to address the issue.

At the most basic level, the root causes of cost overrun and schedule delay may be grouped into three categories, each of which will be considered in turn: (1) bad luck or error; (2) optimism bias; and (3) strategic misrepresentation.

**Error**

Bad luck, or the unfortunate resolution of one of the major project uncertainties mentioned above, is the explanation typically given by management for a poor outcome. The problem with such explanations is that they do not hold up in the face of statistical tests.

Explanations that account for overruns in terms of bad luck or error have been able to survive for decades only because data on project performance has generally been of low quality, i.e. data has been disaggregated and inconsistent, because it came from small-N samples that did not allow rigorous statistical analyses. Once higher-quality data was established that could be consistently compared across projects in numbers high enough to establish statistical significance, explanations in terms of bad luck or error collapsed.

First, if underperformance was truly caused by bad luck and error, we would expect a relatively unbiased distribution of errors in performance around zero. In fact, the data for nuclear projects show with very high statistical significance that the cost overrun distribution does not center on zero (average = 134%, median = 67%, $p < 0.001$) and that the forecasting error is biased towards overrun (cost overrun in 96% of observations, $p < 0.001$).

Second, if bad luck or error were main explanations of underperformance, we would expect an improvement in performance over time, since in a professional setting errors and their sources would be recognized and addressed through the refinement of data, methods, etc., much like in weather forecasting or medical science.

Substantial resources have in fact been spent over several decades on improving data and methods in major project management, including in cost and schedule forecasting. Still the evidence shows that this has not led to improved performance in terms of lower cost overruns and delays.

Bad luck or error, therefore, do not appear to explain the data.

**Optimism bias**

Psychologists tend to explain the underestimation of cost and schedule risks in terms of optimism bias, that is, a cognitive predisposition found with most people to judge future events in a more positive light than is warranted by actual experience. Kahneman and Tversky's (1979a, b) found that human judgment is generally optimistic due to overconfidence and insufficient regard to distributional information about outcomes.

Thus, people will underestimate the costs, completion times, and risks of planned actions, whereas they will overestimate the benefits of the same actions. Similarly, the cost and time needed to complete a project will be optimistic, i.e. under estimated. Such errors of judgment are shared by experts and laypeople alike, according to Kahneman and Tversky.

From the point of view of behavioral science, the mechanisms of scope changes, complex interfaces, archaeology, geology, bad weather, business cycles, etc. are not unknown to planners of capital projects, just as it is not unknown to planners that such mechanisms may be mitigated, for instance by Reference Class Forecasting.

However, planners often underestimate these mechanisms and mitigation measures, due to overconfidence bias, the planning fallacy, and strategic misrepresentation. In behavioral terms, scope changes etc. are manifestations of such underestimation on the part of planners, and it is in this sense that bias and underestimation are the root causes of cost overrun. But because scope changes etc. are more visible than the underlying root causes, they are often mistaken for the cause of cost overrun.

In behavioral terms, the causal chain starts with human bias which leads to underestimation of scope during planning which leads to unaccounted for scope changes during delivery which lead to cost overrun. Scope changes are an intermediate stage in this causal chain through which the root causes manifest themselves.

With behavioral science we say to planners, "*Your biggest risk is you*." It is not scope changes, complexity, etc. in themselves that are the main problem; it is how human beings misconceive and underestimate these phenomena, through overconfidence bias, the planning fallacy, etc. This is a profound and proven insight that behavioral science brings to capital investment planning.

Behavioral science entails a change of perspective: *The problem with cost overrun is not error but bias*, and as long as you try to solve the problem as something it is not (error), you will not solve it. Estimates and decisions need to be de-biased, which is fundamentally different from eliminating error (Kahneman et al. 2011, Flyvbjerg 2008, 2013).

Furthermore, *the problem is not even cost overrun, it is cost underestimation*. Overrun is a consequence of underestimation, with the latter happening upstream from overrun, often years before overruns manifest. Again, if project planners and decision makers try to solve the problem as something it is not (cost and schedule overruns), you will fail. Planners and decision makers need to solve the problem of cost underestimation to solve the problem of cost overrun. Until these basic insights from behavioral science are understood, it is unlikely to get capital investments right, including cost and schedule estimates.

**Political bias**

Economists and political scientists tend to explain underreporting of budget and schedule risks in terms of strategic misrepresentation, or political bias (Wachs 1989, Flyvbjerg 2005). Here, when forecasting the outcomes of projects, forecasters and planners deliberately and strategically overestimate benefits and underestimate cost and schedule in order to increase the likelihood that it is their projects, and not the competition's, that gain approval and funding.

According to this explanation, actors purposely spin scenarios of success and gloss over the potential for failure. This results in managers promoting ventures that are unlikely to come in on budget or on time, or to deliver the promised benefits.

Political bias can be traced to political and organizational pressures, for instance competition for scarce funds or jockeying for position, and to lack of incentive alignment.

The key problem that leads to political bias is a lack of accountability for the parties involved in project development and implementation:

(1) Because of the time frames that apply to major project development and implementation, politicians involved in producing overoptimistic forecasts of project viability in order to have projects approved are often not in office when actual viability can be calculated.
(2) Special interest groups can promote projects at no cost or risk to themselves. Others will be financing the projects, and often taxpayers' money is behind them, including in the form of sovereign guarantees. This encourages rent-seeking behavior for special interest groups.
(3) Contractors, who are an interest group in its own right, are eager to have their proposals accepted during tendering. Contractual penalties for producing over-optimistic tenders are often low compared to the potential profits involved. Therefore, costs and risks are also often underestimated in tenders. The result is that real costs and real risks often do not surface until construction is well under way.

Explanations of cost and schedule overruns in terms of political bias account well for the systematic underestimation of costs and schedule found in the data. A politically biased estimate of costs would be low, resulting in cost overrun, a politically biased estimate of schedule would be short, resulting in delays.

Optimism bias and political bias are both deception, but where the latter is deliberate, the former is not. Optimism bias is self-deception.

**Summary of the root causes**

Research into the track record of past estimates (e.g. Flyvbjerg et al. 2004, Flyvbjerg 2014, 2016) shows that project cost and schedule estimates are systematically and consistently lower than actual outturn cost and actual schedule.

The data show that conventional, inside-view cost and schedule estimates are biased, i.e. they systematically underestimate cost and schedule risks. The data do not fit the "error" explanation of overrun and raise doubts that better models and better data on their own will improve forecasts.

This leaves optimism and political bias as the best explanations of why cost and schedule are underestimated.

As illustrated schematically in Figure 7, explanations in terms of optimism bias have their relative merit in situations where political and organizational pressures are absent or low, whereas such explanations hold less power in situations where political pressures are high.

**Figure 7 Optimism and Political Bias**

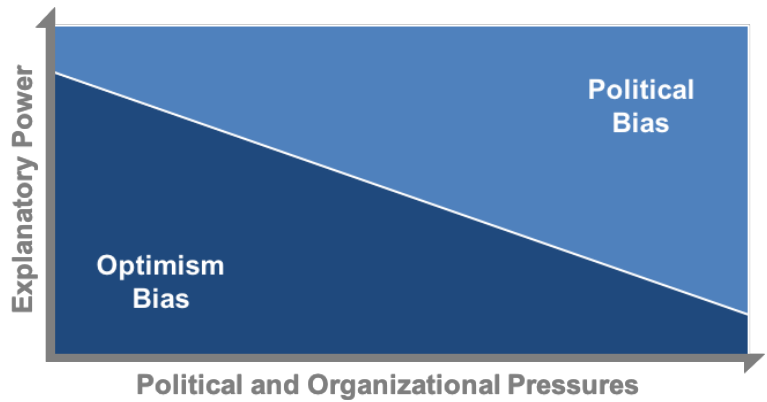

Conversely, explanations in terms of strategic misrepresentation have their relative merit where political and organizational pressures are high, while they become less relevant when such pressures are not present.

Although the two types of explanation are different, the result is the same: inaccurate forecasts and inflated benefit-cost ratios.

Thus, rather than compete, the two types of explanation complement each other: one is strong where the other is weak, and both explanations are necessary to understand the pervasiveness of inaccuracy and risk in project budgeting and scheduling – and how to curb it.